\documentclass[12pt]{article}

\usepackage[margin=1in]{geometry}
\usepackage{setspace}
\usepackage[T1]{fontenc}
\usepackage[utf8]{inputenc}
\usepackage{lmodern}
\usepackage{microtype}
\usepackage{float}

\usepackage{amsmath, amssymb, amsthm, mathtools}
\numberwithin{equation}{section}
\theoremstyle{plain}
\newtheorem{theorem}{Theorem}

\theoremstyle{definition}
\newtheorem{definition}{Definition}
\theoremstyle{remark}
\newtheorem{remark}{Remark} 

\usepackage{graphicx}
\usepackage{booktabs}
\usepackage{caption}
\usepackage{subcaption}

\usepackage[hidelinks]{hyperref}
\usepackage[nameinlink,capitalise,noabbrev]{cleveref}

\usepackage[authoryear,round]{natbib}
\title{Asymptotic linear dependence and ellipse statistics for
multivariate two-sample homogeneity test}

\author{%
  Chifeng Shen$^{a}$,
  Yuejiao Fu$^{a,*}$,
  Michael Chen$^{a}$,
  Xiaoping Shi$^{b}$%
}

\date{}

\begin{document}
\maketitle

\begin{center}
\textit{$^{a}$Department of Mathematics and Statistics, York University, Toronto, ON M3J 1P3, Canada}\\[2pt]
\textit{$^{b}$Department of Computer Science, Mathematics, Physics and Statistics,
University of British Columbia, Kelowna, BC V1V 1V7, Canada}
\end{center}

\vspace{6pt}
\hrule
\vspace{2pt}
\noindent\textit{\footnotesize $^{*}$Correspondence: yuejiao@yorku.ca}
\vspace{10pt}

\begin{abstract}
Statistical depth, which measures the center-outward rank of a given sample with respect to its underlying distribution, has become a popular and powerful tool in nonparametric inference. In this paper, we investigate the use of statistical depth in multivariate two-sample problems. We propose a new depth-based nonparametric two-sample test, which has the $\chi^2_1$ asymptotic distribution under the null hypothesis. Simulations and real-data applications highlight the efficacy and practical value of the proposed test.
\end{abstract}

\noindent\textbf{MSC (2020):} Primary 62G10\\
\noindent\textbf{Keywords:} multivariate two-sample test, statistical depth, elliptical distribution, homogeneity test

\section{Introduction}
The objective of the multivariate two-sample homogeneity test is to determine whether two observed multivariate datasets stem from the same underlying distribution. The alternative hypothesis characterizes discrepancies between distributions, typically in terms of a location shift, a scale change, or a combination of both. The two-sample homogeneity test is crucial in numerous fields and essential for informed decision-making and inference. For instance, 
researchers in behavioral ecology often need to differentiate between the central tendencies of two samples \citep{ecology}. Financial analysts frequently compare financial metrics such as returns, volatility, and correlations of two investment portfolios to evaluate their performance and risk \citep{finance}. In microbiology, researchers analyze differential abundance between two conditions \citep{microbiology}. Two-sample homogeneity tests allow researchers to gain deeper insights into underlying patterns and disparities in data, thereby enhancing decision-making processes.

For univariate data, the classical parametric test is the two-sample $t$-test, which compares the means of two samples drawn from normally distributed populations with equal variance. Hotelling's $T^2$ test extends this approach to multivariate data analysis \citep{hotel1,hotel2}. However, the normality assumption for multivariate data is often challenging to validate. To overcome the limitations of parametric methods, various non-parametric two-sample homogeneity tests have been developed. Cramér test is consistent and invariant with respect to orthogonal linear transformations and sensitive against location shifts \citep{cramertest}. The Energy Distance test proposed by \citet{energytest} measures the discrepancy of underlying distributions based on the Euclidean distance between sample elements, and performs particularly well in high dimension. The connection between the multivariate Wasserstein test and the Energy Distance test was explained by \citet{family}. Additionally, kernel-based two-sample homogeneity tests have been proposed, which use the maximum mean discrepancy (MMD) to detect mean differences \citep{kernel1, maxmean}.

Many robust inference methods have been proposed in the literature, leveraging the natural center-outward ranking provided by multivariate statistical depths \citep{chara1}. \cite{zuo2000general} provided the general notion of depth function and its desirable statistical properties for multivariate data. Based on the concept of data depth and motivated by the quality control problem, \cite{liu1993} proposed a multivariate rank-based test. They proposed a quality index, $\mathcal{Q}(F,G)$, which measures the overall outlyingness of one distribution $G$ with respect to its reference distribution $F$. Let $x_1,x_2,\cdots,x_m$ and $y_1,y_2,\cdots,y_n$ be two independent samples raised  from the distributions $F$ and $G$, respectively. For a depth function $D$, $\mathcal{Q}(F,G) = \int R(y;F) \mathrm{d} G(y),$ where $R(y;F) =\int 1_{\{D(x;F) \leq D(y;F)\}}\mathrm{d}F(x).$ Given the two-sample data, the quality index $\mathcal{Q}(F,G)$ can be estimated by $\mathcal{Q}(F_m,G_n)$, where $F_m$ and $G_n$ are the empirical distributions. Later, \cite{zuo2006limiting}, \cite{shi1} and \cite{variations} proved the conjectured limiting distribution of the Liu-Singh statistic under different sets of regularity conditions. One major drawback of the Liu-Singh statistic lies in the asymmetry in the pair of sample quality indexes, i.e., $\mathcal{Q}(F_m,G_n) \neq \mathcal{Q}(G_n,F_m)$, leading the test result to depend on the choice of the reference distribution. \cite{liu1993} discussed a symmetrized modification by using the pooled sample to calculate the depth value. \cite{liu2006} and \cite{mod3} extended the Liu-Singh test to multi-sample multivariate scale test. They also generalized the Siegel-Tukey \citep{siegel1960nonparametric} and Ansari-Bradley  \citep{ansari1960rank} scale tests for the multivariate data setting. Recently, not limiting to the scale test, several new statistics have been proposed by combining pairwise Liu-Singh statistics to restore symmetry in the two samples. For example, \cite{shi1} proposed the Weighted average statistic $W_{m,n}(\omega)$ and the Maximum statistic $M_{m,n}$. \cite{variations} also investigated extensively variations of the Liu-Singh statistic with clear geometric illustrations and extended the tests from multivariate to functional data.

\begin{figure}[H]
    \centering    \includegraphics[width=0.41\linewidth]{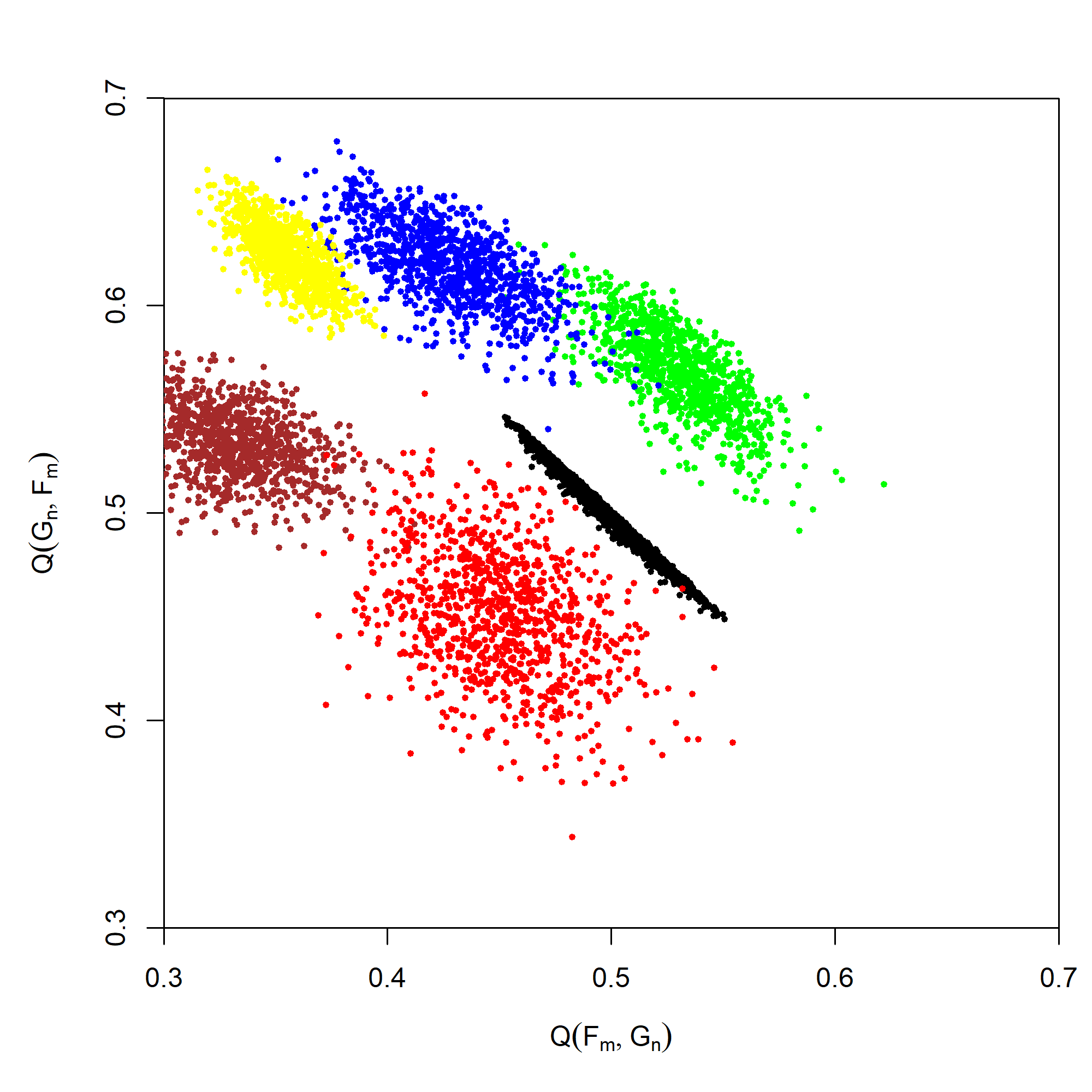}    
    \caption{\textbf{Black Points}: $F=G$; \textbf{Points in other color}: $F\neq G$. The number of replications is 1000, and $m=n=800$.The pairs $(\mathcal{Q}(F_m, G_n), \mathcal{Q}(G_n, F_m))$ represent the Liu–Singh statistics.}
    \label{fig:1}
\end{figure}

Inspired by the work of \cite{shi1} and \cite{variations}, we propose a new statistic that combines the pair of Liu-Singh statistics, boosting the power of the two-sample homogeneity test. This statistic leverages the complementary strengths of the pair of Liu-Singh statistics to deliver more accurate and reliable results. The main idea is to design the non-rejection region of the proposed test to enhance its power while ensuring the significance level remains controlled, leveraging information from the null limiting distribution of the test statistic. As illustrated by an example in Figure \ref{fig:1}, the pair of statistics represents the Liu–Singh statistics computed using Mahalanobis depth. The specific distributions used are detailed in the Appendix \ref{appb}. Under the alternative hypothesis, the generated points (colored) may appear anywhere around those generated under the null hypothesis (black). Therefore, it is crucial for the non-rejection region to be smaller, well-defined, and closed, while adequately covering enough points generated under the null hypothesis. More broadly, this approach can be extended to pairs of statistics that are asymptotically linearly dependent, and elliptical (see Def \ref{def1}). 
\begin{definition}\label{def1}
    Let $\mathbf{x}_m=(x_1,x_2,\cdots,x_m) \in \mathbb{X}$ and $\mathbf{y}_n=(y_1,y_2,\cdots,y_n) \in \mathbb{Y}$ be the samples drawn from distribution $F$ and $G$, respectively. Consider two statistics $h_1 (\mathbf{x}_m,\mathbf{y}_n) $ and $h_2(\mathbf{x}_m,\mathbf{y}_n) .$ 
    For $m,n \to \infty,$
    $
    \frac{m}{m+n} \to \tau$ for some  $0 < \tau <1,
    $
    if
    $$
    \sqrt{\frac{mn}{m+n}}\begin{pmatrix}
    h_1(\mathbf{x}_m,\mathbf{y}_n)-a\\
    h_2(\mathbf{x}_m,\mathbf{y}_n)-b
    \end{pmatrix} \xrightarrow{\text{d}} \mathbf{z},
    $$
    where $a, b\in \mathbf{R}$,  $\mathbf{z} \sim \mathbf{MVN}\Big(\begin{pmatrix}
    0\\
    0
    \end{pmatrix}, \begin{pmatrix}
    \sigma^2_1&\rho\sigma_1\sigma_2\\
    \rho\sigma_1\sigma_2&\sigma^2_2
    \end{pmatrix}\Big)$, $\rho \in [-1,0)\cup(0,1]$, and $\sigma_1,\sigma_2 > 0.$
    Then, the two statistics $h_1(\mathbf{x}_m,\mathbf{y}_n)$ and $h_2(\mathbf{x}_m,\mathbf{y}_n)$ are called asymptotically linearly dependent, and elliptical (ALDE).
\end{definition}

\begin{remark}
    When $\rho=0$, the pair of statistics is asymptotically linearly independent. The scatter plot of the pairs of statistics forms a standard ellipse with a shifted center.
\end{remark} 

\begin{definition}\label{def2} Let $m,n \in \mathbf{Z}^+$, $\theta \in [-\frac{\pi}{2},\frac{\pi}{2}]$, and $C,\lambda >0.$
\begin{equation*}
    ER(m,n,\theta,\lambda,C)=\Big\{(u,v)|\frac{mn}{m+n}\begin{pmatrix}
     u\\
     v
\end{pmatrix}^T\begin{pmatrix}\cos^2{\theta}+\lambda \sin^2{\theta}& (1-\lambda)\sin{\theta}\cos{\theta}\\
    (1-\lambda)\sin{\theta}\cos{\theta} & \sin^2{\theta}+\lambda \cos^2{\theta}
    \end{pmatrix}
    \begin{pmatrix}
     u\\
     v
\end{pmatrix}\leq C\Big\}    
\end{equation*}
is called an elliptical region.
\end{definition}

\begin{figure}[h]
    \centering
    \includegraphics[width=0.6\linewidth]{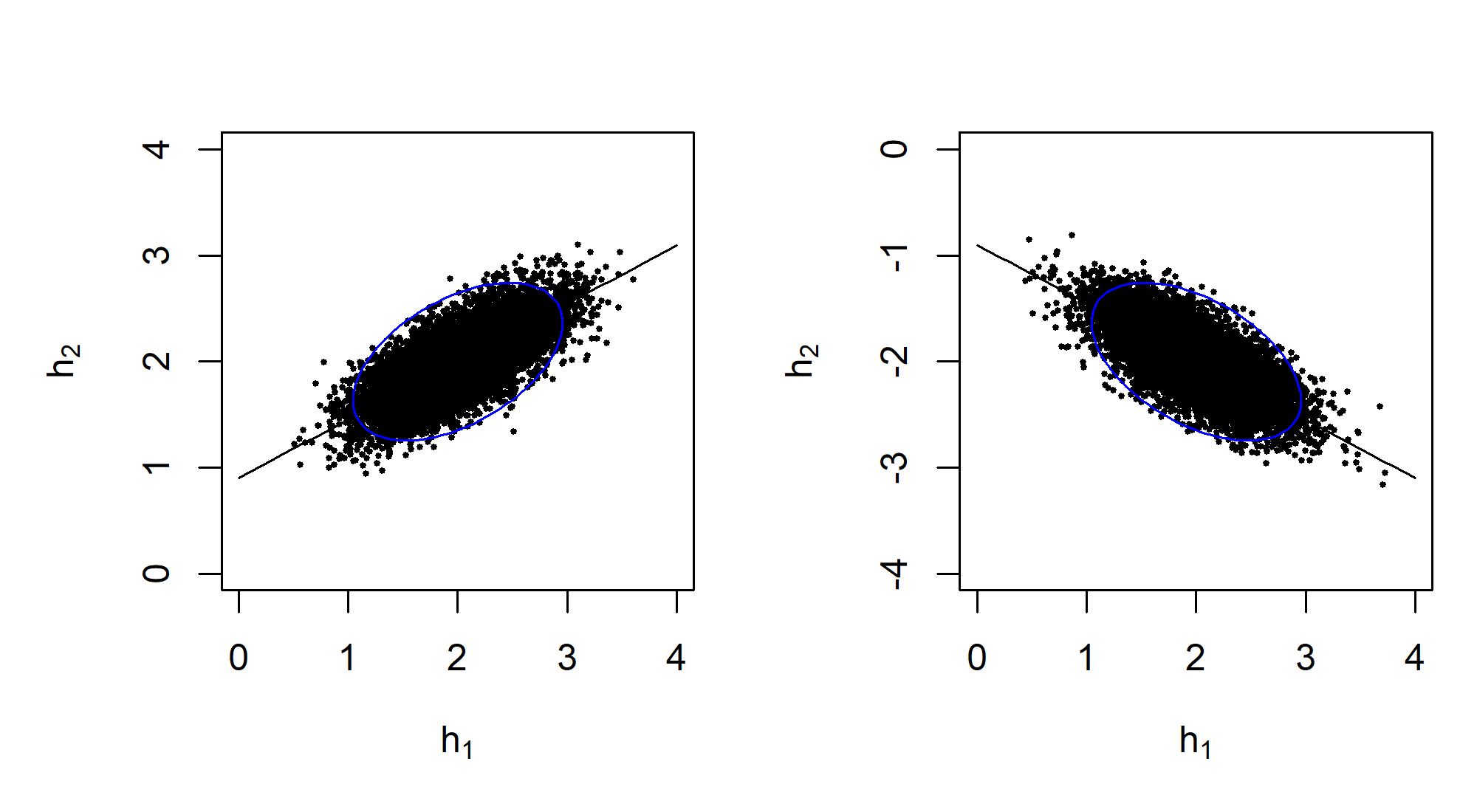}    
    \caption{Left plot:  \textbf{Case I}; Right plot: \textbf{Case II}. The number of replications is 1000, and $m=n=50$. The blue circle is the ellipse with $\alpha=0.05,$ and $\lambda=0.3.$}
    \label{fig:2}
\end{figure}

For the pair of ALDE statistics $(h_1(\mathbf{x}_m,\mathbf{y}_n),     h_2(\mathbf{x}_m,\mathbf{y}_n))$, we propose the elliptical non-rejection region with the following parameter specifications. The ratio of the minor axis to the major axis of the ellipse, $\lambda$, depends on the prior information. The counterclockwise rotation angle, $\theta$, equals $\frac{\pi}{2}+\arctan\frac{\nu_2}{\nu_1}$, where $(\nu_1,\nu_2)^T$ is the eigenvector corresponding to the largest eigenvalue of the variance-covariance matrix. The value of $C$ depends on the limiting distribution, and the significance level. This can be visualized in Figure \ref{fig:2} based on cases I and II as follows.

Let $F=G \sim N(2,3)$, and $h_1(\mathbf{x}_m,\mathbf{y}_n)=\frac{1}{m}\sum_{i=1}^m x_i.$

\textbf{Case I}: Let 
$h_2(\mathbf{x}_m,\mathbf{y}_n)=\frac{1}{m+n}\sum_{i=1}^m\sum_{j=1}^n (x_i+y_j).$

\textbf{Case II}: Let 
$h_2(\mathbf{x}_m,\mathbf{y}_n)=-\frac{1}{m+n}\sum_{i=1}^m\sum_{j=1}^n (x_i+y_j).$

The paper is outlined as follows. In Section 2, we propose the new multivariate two-sample homogeneity test statistic and establish its null limiting distribution. The motivation behind the new test is further elucidated through geometric illustrations, providing a visual understanding of the conceptual framework. In Section 3, simulation studies demonstrate the good performance of the proposed test using three different depth functions: Mahalanobis depth \citep{liu1993}, Spatial depth \citep{spatial1,spatial2, spatial3}, and Projection depth \citep{liu1992data,zuo2000general}. Two real data examples are shown in Section 4. Conclusions and future work are presented in Section 5.

\section{Main Results}
Consider two $d$-dimensional samples $x_1,x_2,\cdots,x_m$ and $y_1,y_2,\cdots,y_n$ randomly and independently raised  from the distributions $F$ and $G$, respectively. For a depth function $D$, \cite{liu1993} introduced the relative deepness of a point $y$ with respect to the probability measure $F$ as follows,
\begin{equation*}
    R(y;F) =\int 1_{\{D(x;F) \leq D(y;F)\}}\mathrm{d}F(x).
\end{equation*}
With the reference distribution $F$, the Liu-Singh statistic is defined as
\begin{equation*}
    \mathcal{Q}(F,G) = \int R(y;F) \mathrm{d} G(y).
\end{equation*}

Without the information about the distributions $F$ and $G$, the empirical distributions $F_m$ and $G_n$ can be used for the following sample version,
\begin{equation*}
    R(y_j;F_m) = \frac{1}{m}\sum_{i=1}^{m}1_{\{D(x_i;F_m) \leq D(y_j;F_m)\}},
\end{equation*}
\begin{equation*}
   \mathcal{Q}(F_m,G_n) =\frac{1}{n} \sum_{j=1}^{n} R(y_j;F_m). 
\end{equation*}
In general, the $\mathcal{Q}$-statistics are asymmetric, leading to a different test decision when the reference distributions are exchanged. To tackle the drawback, \cite{shi1} introduced the weighted average statistic $W_{m,n}(\omega)$ and maximum statistic $M_{m,n}$ as follows,
\begin{equation}
    W_{m,n}(\omega)= \frac{12mn}{m+n}\big[\omega\big(\mathcal{Q}(F_m,G_n)-\frac{1}{2}\big)^2+(1-\omega)\big(\mathcal{Q}(G_n,F_m)-\frac{1}{2}\big)^2\big],
\end{equation}
where $0\leq \omega \leq 1.$
\begin{equation}
    M_{m,n}=\frac{12mn}{m+n} \max \{\big(\mathcal{Q}(F_m,G_n)-\frac{1}{2}\big)^2,\big(\mathcal{Q}(G_n,F_m)-\frac{1}{2}\big)^2\}.
\end{equation}
There are various statistical depth functions that can be applied for the new test. In this paper, we focus on Mahalanobis depth, Spatial depth, and Projection depth, which are reviewed below.

For any observation $x \in \mathcal{R}^d$ and a given $d-$dimensional distribution $F$, the Mahalanobis depth \citep{liu1993} of $x$ with respect to the distribution $F$ is defined as follows,
\begin{equation*}
   MD(x;F)=\frac{1}{1+(x-\mu_F)^{T}\Sigma_F^{-1}(x-\mu_F)}, 
\end{equation*}
where $\mu_F$ and $\Sigma_F$ are the mean and variance-covariance matrix of $F.$ The Spatial depth (also $L_1-$depth) \citep{spatial1,spatial2, spatial3} of $x$ with respect to the distribution $F$ is defined as follows,
\begin{equation*}
   SD(x;F) = 1-||E_X\frac{(x-X)}{||x-X||}||,
\end{equation*}
where $X \sim F.$ The projection depth \citep{liu1992data,zuo2000general} is defined as
\begin{equation*}
   PD(x;F)=\frac{1}{1+O(x;F)},
\end{equation*}
where $O(x;F)=\sup_{||\nu||=1}\frac{|\nu^Tx-Med(\nu^TX)|}{Med|\nu^TX-Med(\nu^Tx)|},$ $X \sim F,$ and $Med$ is the median function.

Given a $d$ dimensions dataset $\mathbf{x}=\{x_1,x_2,\cdots,x_m\}$ and a point $x$ in $\mathcal{R}^d$, the sample versions for these depths are adapted as follows. 
\begin{itemize}
    \item Mahalanobis Depth
    \begin{equation*}
    MD(x;F_m)=\sqrt{(x-\Bar{x})^TS^{-1}(x-\Bar{x})},
    \end{equation*}
    where $\Bar{x}$ is the sample mean vector, and $S$ is the sample variance-covariance matrix matrix.
    \item Spatial Depth
    \begin{equation*}
    SD(x;F_m)=1-||\frac{1}{m}\sum_{i=1}^m\frac{(x-x_i)}{||x-x_i||}||.
    \end{equation*}
    \item Projection Depth
    \begin{equation*}
    PD(x;F_m) = \frac{1}{1+O(x;F_m)},
    \end{equation*}
    where
    \begin{equation*}
    O(x;F_m)=\sup_{||\nu||=1}\frac{|\nu^Tx-Med_{1\leq i \leq m}(\nu^Tx_i)|}{Med_{1 \leq i \leq m}(|\nu^Tx_i-Med_{1 \leq j \leq m}(\nu^Tx_j)|)}
    \end{equation*}
\end{itemize}

For a given depth function $D(\cdot;\cdot)$ with $0\leq D(x;H) \leq 1$, \citep{variations} proposed four conditions on depths as follows,
\begin{enumerate}
    \item [(A1)]
\emph{$\mathbb{P}(D(X, F) \in [y_1, y_2]) \leq C |y_2 - y_1|^{\beta}$ for positive constants $C, \beta$ and any $y_1, y_2 \in \mathbb{R}_{\geq 0}$, for some $1/2 < \beta \leq 1$.}
    \item [(A2)]
\emph{$
\mathbb{E} \left( \sup_{x \in \Omega} \left| D(x, F_m) - D(x, F) \right|^{2\beta} \right) = O(m^{-\beta})
\quad \text{with } 1/2 < \beta \leq 1 \text{ as in } \text{A}{1}.
$}
    \item [(A3)]
\emph{There exist a \textit{deterministic} constant $C_{\text{det}}$ and an index $m_0$ such that for every $m \geq m_0$:
$
\sup_{x \in \Omega} \left| D(x, F_m) - D(x, F^{-\{1\}}_m) \right| \leq \frac{C_{\text{det}}}{m} \quad \text{almost surely,}
$ where $F^{-I}_m$ for some $I \in \{1, \ldots, n\}$ is the empirical probability measure with respect to $\{X_i : i \in \{1, \ldots, n\} \setminus I\}$.}
    \item [(A4)]
\emph{Let $X$ be an independent copy of $X_1, \ldots, X_m$. Then, for any constant $C > 0$, it holds}
\[
\mathbb{E} \left( \left[ \mathbb{P} \left( \left| D(X_1, F^{-\{1\}}_m) - D(X, F^{-\{1\}}_m) \right| \leq \frac{C}{m} \,\middle|\, X_2, \ldots, X_m \right) \right]^2 \right)
= O(m^{-2\beta})
\]
\emph{with $1/2 < \beta \leq 1$ as in \text{A}{1}.}
\end{enumerate} 

\cite{zuo2006limiting} showed that, under their regularity conditions, $\mathcal{Q}(F_m,G_n)-\frac{1}{2}=\frac{1}{2}-\mathcal{Q}(G_n,F_m)+o_p(n^{-1/2})+o_p(m^{-1/2})$ . Under the same conditions, \citep{shi1} established that the asymptotic distribution of both statistics $W_{m,n}(\omega)$ and $M_{m,n}$ is $\chi^2_1$. The result also holds under the assumptions A1-A4 from \cite{variations}.

\begin{figure}[h]
    \centering    \includegraphics[width=0.75\linewidth]{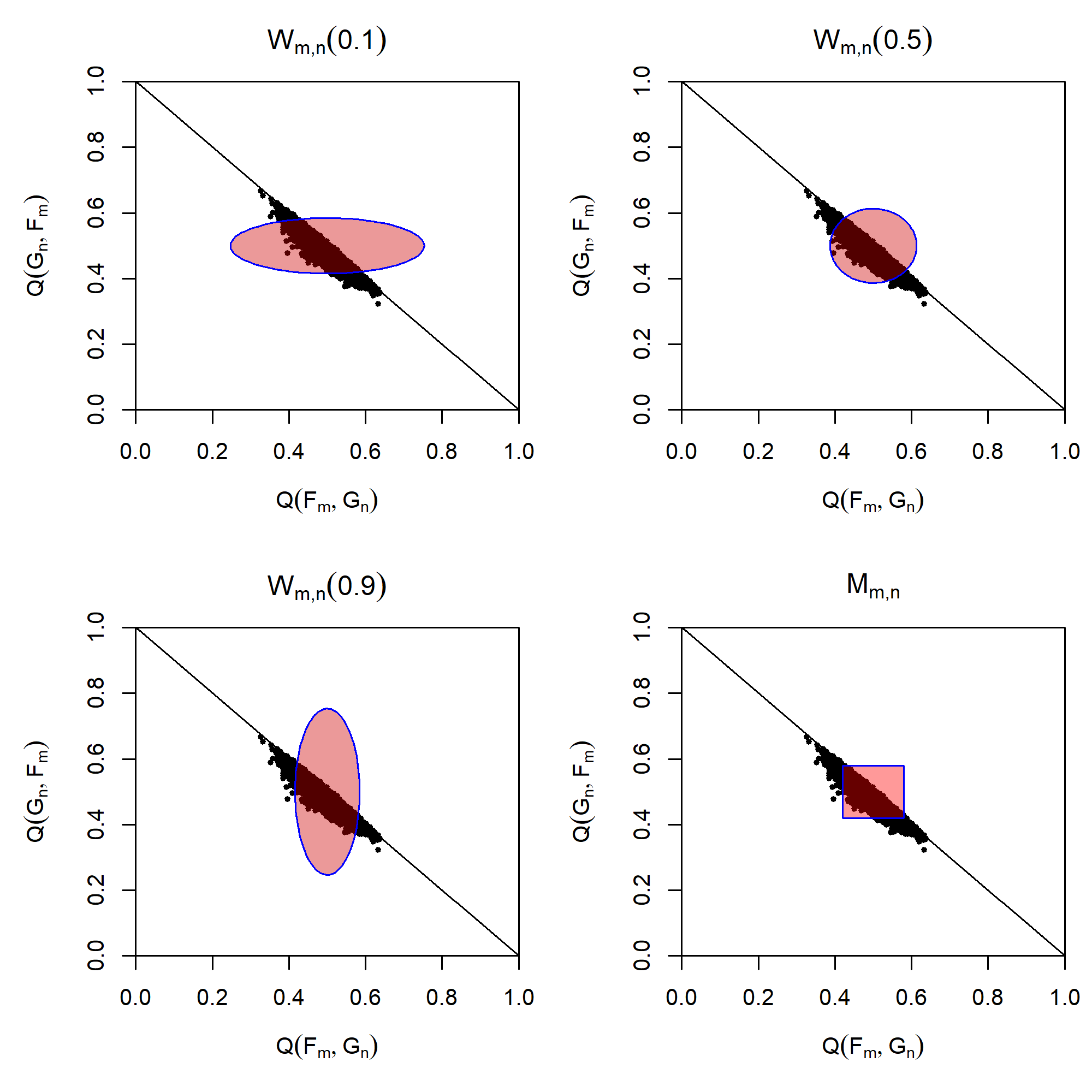}    
    \caption{The non-rejection regions of $W_{m,n}$ for $\omega=0.1,0.5,0.9$ and $M_{m,n}$, when $\alpha=0.05.$ The black points are 1,000 $(\mathcal{Q}(F_m,G_n),\mathcal{Q}(G_n,F_m))$ computed using Mahalanobis depth under the null hypothesis, $F=G=N(\Vec{0},\mathbf{I}_{2}).$}
    \label{fig:3}
\end{figure}

The geographic illustration shows the shape of the non-rejection region \citep{variations}. When $\alpha=0.05$ and $m=n=300,$ Figure \ref{fig:3} plots non-rejection regions of the weighted average statistics $W_{m,n}(\omega)$ and maximum statistic $M_{m,n}$, covering 1,000 points of $(\mathcal{Q}(F_m,G_n),\mathcal{Q}(G_n,F_m))$ computed using Mahalanobis depth under the null hypothesis, $F=G=N(\Vec{0},\mathbf{I_2}).$   

In Figure \ref{fig:3}, the non-rejection region of $W_{m,n}(\omega)$ is a standard ellipse, and the region of $M_{m,n}$ is a square. Both have the center $(\frac{1}{2},\frac{1}{2}).$ The points of $(\mathcal{Q}(F_m,G_n),\mathcal{Q}(G_n,F_m))$ tend to centralize to the point $(\frac{1}{2},\frac{1}{2})$ along the line $\mathcal{Q}(F_m,G_n)+\mathcal{Q}(G_n,F_m)-1=0$ as $\min(m,n) \to \infty.$ To increase the power with controlling type I error, it is reasonable to create a non-rejection region which is symmetric about the point $(\frac{1}{2},\frac{1}{2})$ and spread along the line $\mathcal{Q}(F_m,G_n)+\mathcal{Q}(G_n,F_m)-1=0$. Thus, the ellipse statistic $\mathcal{R}(\lambda,\theta,F_m,G_n)$ is conducted as follows, 
\begin{equation}\label{re}
\mathcal{R}(\lambda,\theta,F_m,G_n)=
\frac{12mn}{m+n}
\vec{\mathcal{Q}}^T
\begin{pmatrix}
    \cos^2{\theta}+\lambda \sin^2{\theta}& (1-\lambda)\sin{\theta}\cos{\theta}\\
    (1-\lambda)\sin{\theta}\cos{\theta} & \sin^2{\theta}+\lambda \cos^2{\theta}
\end{pmatrix}
\vec{\mathcal{Q}}
\end{equation}
where $\vec{\mathcal{Q}}=\begin{pmatrix}
    \mathcal{Q}(F_m,G_n)-\frac{1}{2}\\
    \mathcal{Q}(G_n,F_m)-\frac{1}{2}
\end{pmatrix}$,
$\lambda$ is the ratio of minor axis to major axis of the ellipse, and $\theta$ is the counterclockwise rotation angle. 

Under conditions A1-A4 and the null hypothesis, we show that the proposed test statistic has a simple $\big[1+\lambda-(1-\lambda)\sin{2\theta}\big]\chi^2_1$ asymptotic distribution, as stated in the following theorem.\\
\begin{theorem}\label{thm1}
Given two $d$-dimensional random samples $x_1,x_2,\cdots,x_m$ and $y_1,y_2,\cdots,y_n$ raised independently from the distributions $F$ and $G$, respectively. Let $F_m$ and $G_n$ be the corresponding empirical distributions. Let A1-A4 hold true and, for $m,n \to \infty,$
\begin{equation*}
    \frac{m}{m+n} \to \tau \textit{ for some } 0 < \tau <1.
\end{equation*}
Under the null hypothesis $H_0:F=G$,and $\min(m,n) \to \infty$, we have $\mathcal{R}(\lambda,\theta,F_m,G_n) \xrightarrow{\text{d}} \big[1+\lambda-(1-\lambda)\sin{2\theta}\big]\chi^2_1$.
\end{theorem}
The Proof of the theorem \ref{thm1} is presented in Appendix \ref{appa}.

\begin{remark}
$W_{m,n}(\omega)$ is a special case of $\frac{1}{1+\lambda-(1-\lambda)\sin{2\theta}}\mathcal{R}(\lambda,\theta,F_m,G_n)$. They are equivalent when $\theta=0$ and $\lambda=\frac{1-\omega}{\omega}.$    
\end{remark} 

\begin{remark}
It is easy to show that under the null hypothesis, $F=G$, with the critical value from the asymptotic distribution, the intersection points between the boundary of the non-rejection region and the line are $\mathcal{Q}(F_m,G_n)+\mathcal{Q}(G_n,F_m)-1=0$ are  {\small $\begin{cases} 
    \mathcal{Q}(F_m,G_n)=\frac{1}{2} \pm \sqrt{\frac{(m+n)\chi^2_{1-\alpha}(1)}{12mn}}\\
    \mathcal{Q}(G_n,F_m)=\frac{1}{2} \mp \sqrt{\frac{(m+n)\chi^2_{1-\alpha}(1)}{12mn}}
\end{cases}$} for $W_{m,n}(\omega)$, $M_{m,n}$, and $\mathcal{R}(\lambda,\theta,F_m,G_n)$. In other words, the $\lambda$ supply a flexibility to adjust the shape of non-rejection region created by $\mathcal{R}(\lambda,\theta,F_m,G_n)$.
\end{remark}

\begin{figure}[h]
    \centering
    \includegraphics[width=0.8\linewidth]{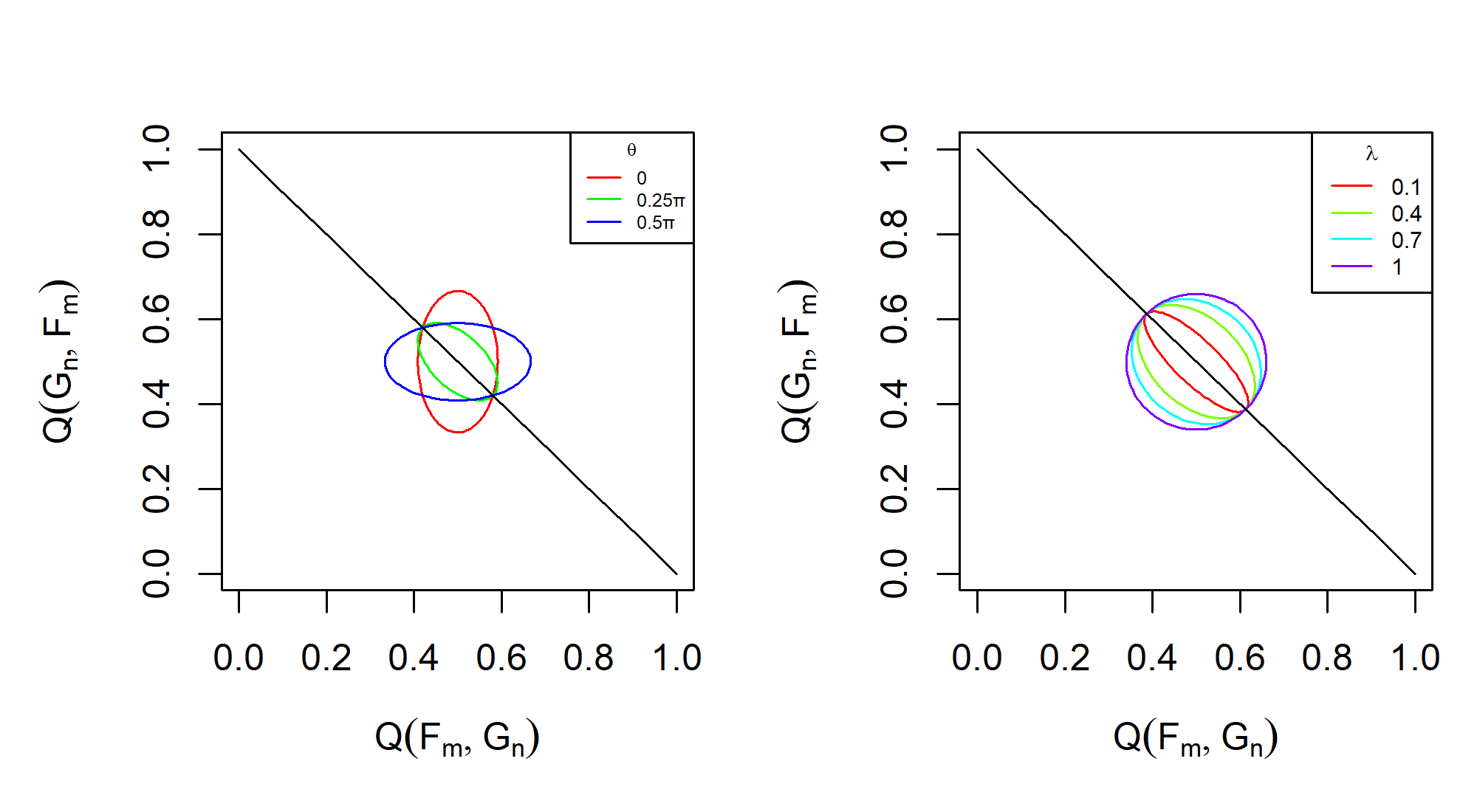}    
    \caption{ The non-rejection region of rotated ellipse statistic $\mathcal{R}(\lambda,\theta,F_m,G_n)$ for $\lambda=0.1,0.4,0.7,1$ with $\theta=\frac{\pi}{4}$  and for $\theta=0,\frac{\pi}{4},\frac{1}{2}\pi$ with $\lambda=0.3$, when $\alpha=0.05.$}
    \label{fig:4}
\end{figure}

Figure \ref{fig:4} shows non-rejection regions of $\mathcal{R}(\lambda,\theta,F_m,G_n)$ for different values of $\lambda$ and $\theta$ with $\alpha=0.05.$ When $\theta=\frac{\pi}{4},$ one axis of the ellipse lies along the line $\mathcal{Q}(F_m,G_n)+\mathcal{Q}(G_n,F_m)-1=0$, and it is symmetric about the point $(\frac{1}{2},\frac{1}{2}).$ Moreover, with a fixed $\theta$, the ellipse becomes more flatten along one axis as the value of $\lambda$ decreasing. It is nature to set $\theta=\frac{\pi}{4}$ and then choose an appropriate $\lambda$ to control the area of ellipse. The Equation \ref{re} can be modified as follows, 
\begin{eqnarray*}
    \mathcal{E}_{m,n}(\lambda)
    &=&
    \mathcal{R}_{m,n}(\lambda,\frac{\pi}{4},F_m,G_n)\\
    &=& \frac{6mn}{m+n}\big[(1+\lambda)(\mathcal{Q}(F_m,G_n)-\frac{1}{2})^2+(1+\lambda)(\mathcal{Q}(G_n,F_m)-\frac{1}{2})^2\\
    &+& 2(1-\lambda)(\mathcal{Q}(F_m,G_n)-\frac{1}{2})(\mathcal{Q}(G_n,F_m)-\frac{1}{2})\big]
\end{eqnarray*}
According to the Theorem \ref{thm1}, $\frac{1}{2\lambda}\mathcal{E}_{m,n}(\lambda)\xrightarrow{\enskip d \enskip} \chi^2_1.$

\begin{figure}[h]
    \centering
    \includegraphics[width=0.8\linewidth]{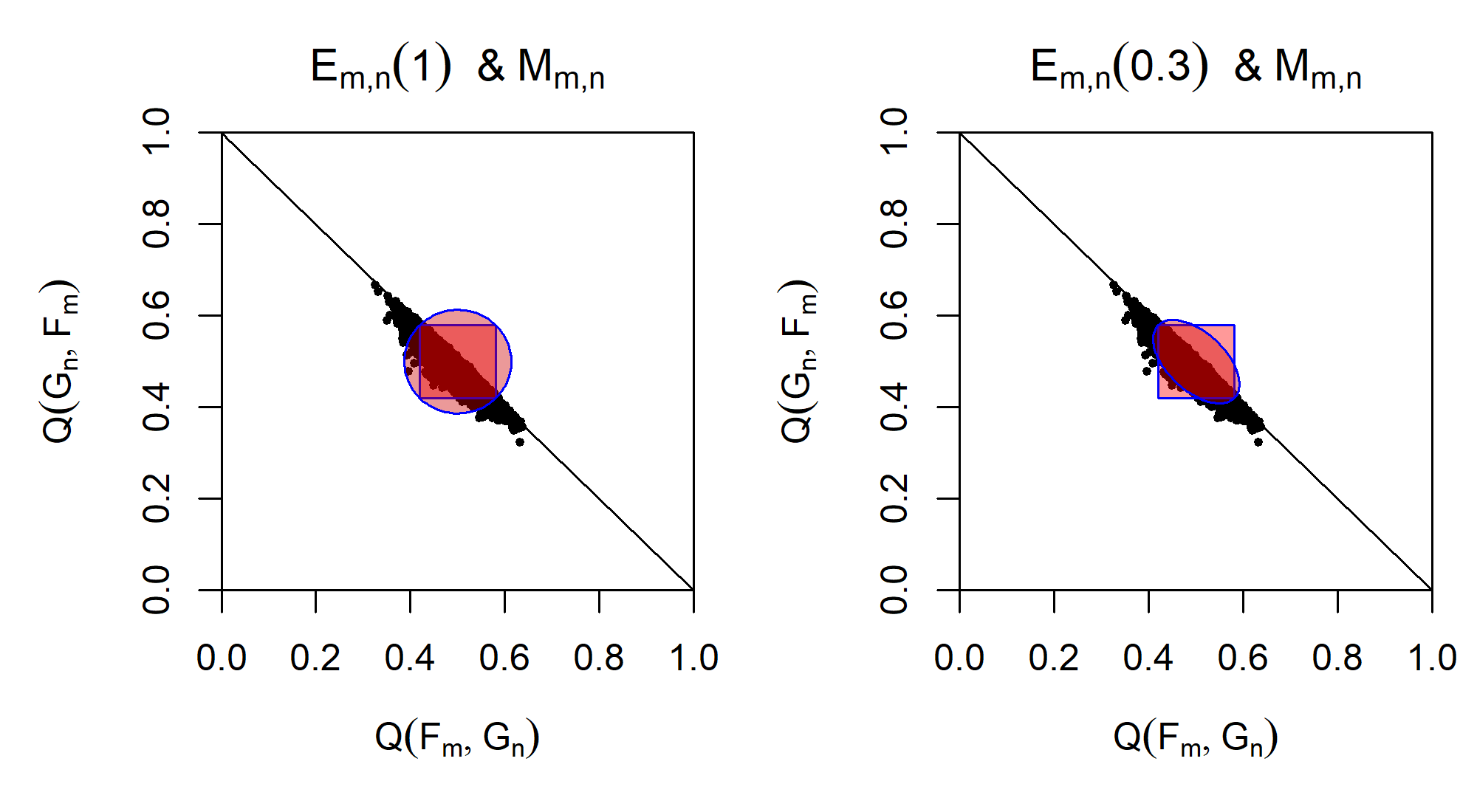}
    \caption{The non-rejection region for $M_{m,n}$ and $ \mathcal{E}_{m,n}(\lambda)$ for $\lambda=0.3,1$, when $\alpha=0.05.$ The black points are 1,000 $(\mathcal{Q}(F_m,G_n),\mathcal{Q}(G_n,F_m))$ computed using Mahalanobis depth under the null hypothesis, $F=G=N(\Vec{0},\mathbf{I_2}).$}
    \label{fig:5}
\end{figure}
Figure \ref{fig:5} compares the non-rejection region for $M_{m,n}$ and $\mathcal{E}_{m,n}$ for $\lambda=0.3,1$, when $\alpha=0.05$ and $m=n=300.$ The black points are 1,000 $(\mathcal{Q}(F_m,G_n),\mathcal{Q}(G_n,F_m))$ computed using Mahalanobis depth under the null hypothesis, $F=G=N(\Vec{0},\mathbf{I_2}).$
Clearly, the ellipse can cover more points with a smaller area than the square. In other words, compared with square, the ellipse improves the power of the proposed test and controls the type I error simultaneously. In the next section, the simulations are conducted to compare the performances between different statistical tests.

\section{Simulation Studies}
The simulation studies are conducted to compare the performance of the proposed tests in the finite two-sample problem. Two random samples $x_1, x_2, \cdots,x_m$ and $y_1,y_2, \cdots, y_n$ are randomly drawn from distributions $F$ and $G$, respectively. Compared with other two-sample homogeneity tests such as MANOVA, Energy, and Wasserstein, $W_{m,n}(\omega)$ and $M_{m,n}$ have better performance \citep{shi1}. Therefore, we examine the empirical size and power of $M_{m,n}$, $W_{m,n}(\omega)$ and $ \mathcal{E}_{m,n}(\lambda)$  using Mahalanobis depth, spatial depth, and projection depth, with $\lambda=0.1,0.3,0.5,0.7,1$. To estimate the empirical power, the critical value from the asymptotic distribution is replaced by a value derived from the empirical quantile of simulated observations under the null hypothesis. For the sample size, we consider $n=m$ and $n=\frac{1}{2}m$, where $m=100,200,\cdots,1000$.

The Type I error is considered at first. Assuming $F=G$ follow a bivariate normal distribution, $N(\Vec{0},\mathbf{I})$, with mean $\Vec{0}$ and identity covariance matrix $I_{2\times2}$. Since $W_{m,n}(0.5)$ and $\mathcal{E}_{m,n}(1)$ are equivalent, we compare $\alpha=0.05$ with the empirical size of the test statistics, $\mathcal{E}_{m,n}(\lambda)$ for $\lambda=0.1,0.3,0.5,0.7,1$, $M_{m,n}$ and $W_{m,n}(\frac{n}{m+n})$ with 10,000 repetitions.

\begin{figure}[h]
    \centering
    \includegraphics[width=0.8\linewidth]{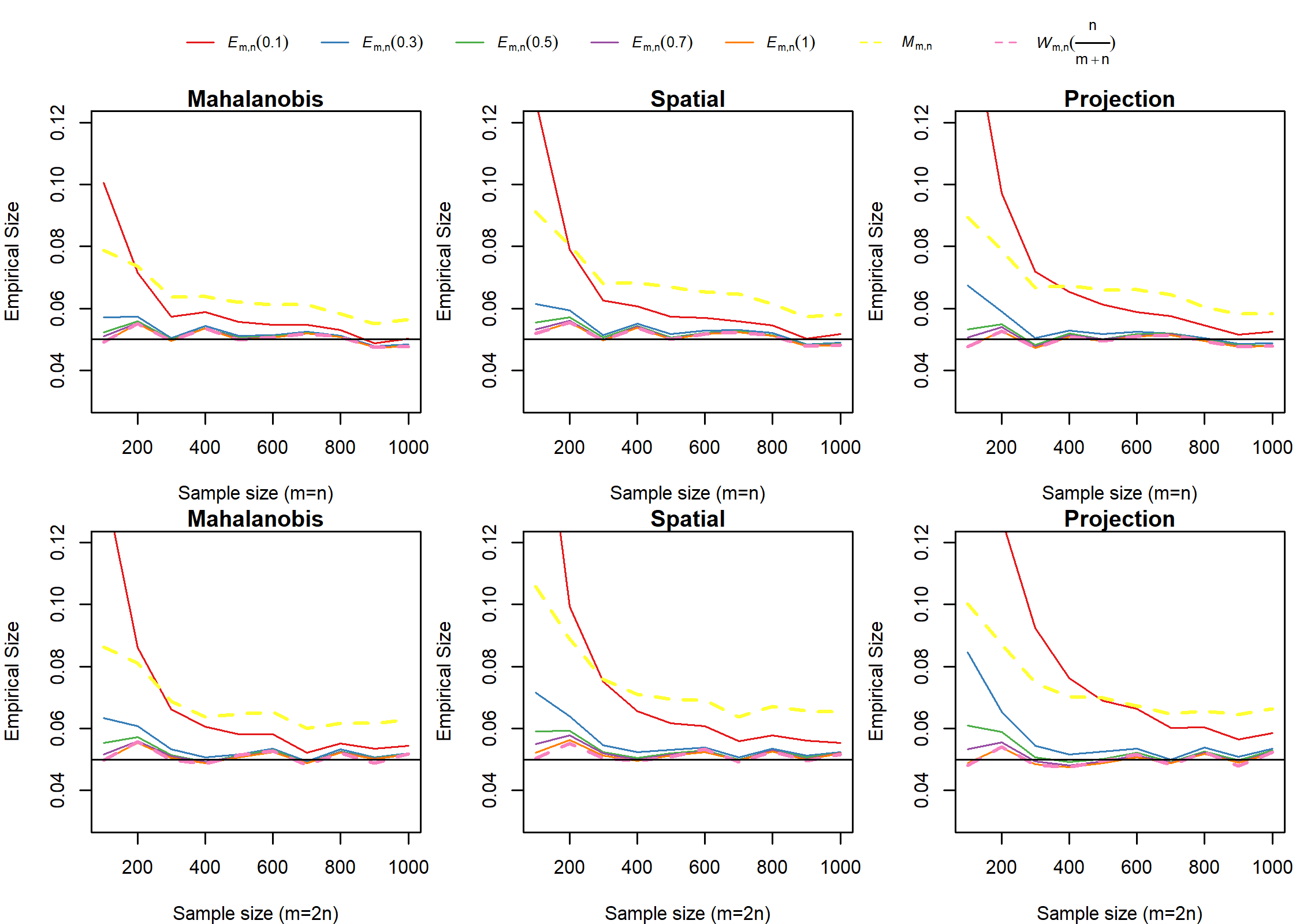}
    \caption{Comparison of empirical size of three statistics, $\mathcal{E}_{m,n}(\lambda)$ and $M_{m,n}$ for $m=100,200, \cdots, 1000$ and $n=m$ (1st row) or $m=2n$ (2nd row). Three depth functions are adopted: Mahalanobis depth (1st column), Spatial depth (2nd column), and Projection depth (3rd column). $\lambda=0.1,0.3,0.5,0.7,1$. }
    \label{fig:6}
\end{figure}

Figure \ref{fig:6} shows the comparisons of empirical size and the theoretical Type I error. For $\mathcal{E}_{m,n}(\lambda)$, the empirical size is closer to $\alpha=0.05$ as increasing the value of $\lambda$ for small sample size, but the difference becomes smaller as $\lambda \geq 0.3$. Compared with $M_{m,n}$,  $\mathcal{E}_{m,n}(\lambda)$ can approach the theoretical Type I error with smaller sample size. Overall, the empirical size of $\mathcal{E}_{m,n}(\lambda)$ is closer to $\alpha=0.05$ for all three depth functions, except $\lambda< 0.3$ under small sample sizes.
 
Next, we consider the empirical power for these statistics: $M_{m,n}$ and $\mathcal{E}_{m,n}(\lambda)$ with $\lambda=0.1,0.3,0.5,0.7,1$ We also consider the sample sizes $n=m$ and $n=\frac{1}{2}m$, where $m=100,200,\cdots,1000$. The empirical power is studied for $\alpha=0.05$ with 1,000 repetitions under the following three alternative hypotheses.

\begin{enumerate}
    \item Two bivariate normal distribution with a scale change, $H_a: F \sim N(\Vec{0},(\begin{smallmatrix}
        1&0\\
        0&1
    \end{smallmatrix}))$ vs $  G \sim N(\Vec{0},(\begin{smallmatrix}
        1&0.5\\
        0.5&1
    \end{smallmatrix}))$. 
    \item Two bivariate normal distributions with a mean change,
    $H_a: F \sim N(\Vec{0},(\begin{smallmatrix}
        1&0\\
        0&1
    \end{smallmatrix}))$ vs $  G \sim N((\begin{smallmatrix}
        0.35\\0.35
    \end{smallmatrix}),(\begin{smallmatrix}
        1&0\\
        0&1
    \end{smallmatrix}))$. 
    \item
    Two bivariate normal distributions with the scale and mean change, $H_a: F \sim N(\Vec{0},(\begin{smallmatrix}
        1&0\\
        0&1
    \end{smallmatrix}))$ vs $ G \sim N((\begin{smallmatrix}
        0.3\\0.3
    \end{smallmatrix}),(\begin{smallmatrix}
        1&0.4\\
        0.4&1
    \end{smallmatrix}))$. 
\end{enumerate}

\begin{figure}[H]
    \centering    \includegraphics[width=0.7\linewidth]{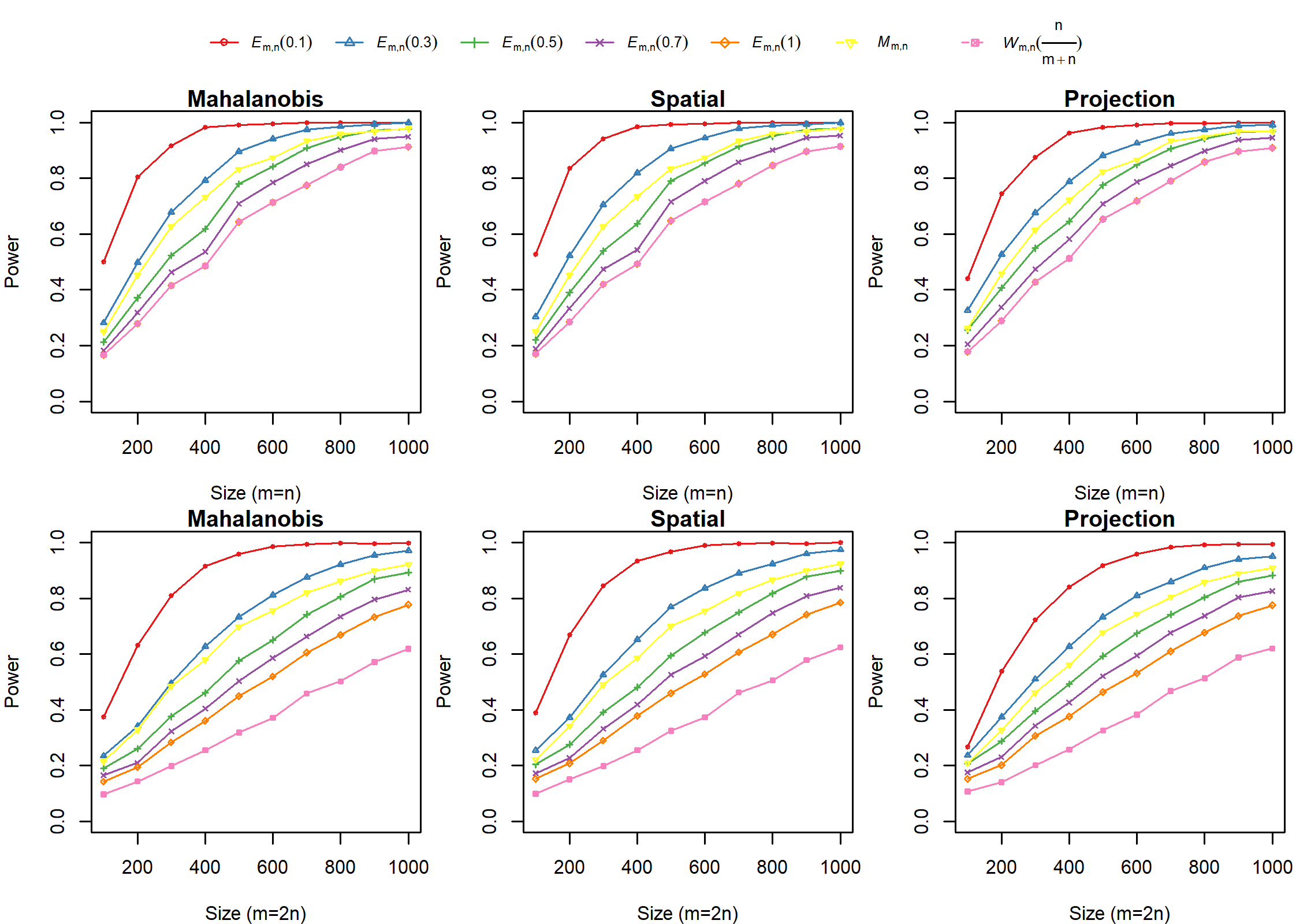}
    \caption{Power comparison under alternative hypothesis $H_a: F \sim N(\Vec{0},(\begin{smallmatrix}
        1&0\\
        0&1
    \end{smallmatrix}))$ vs $  G \sim N(\Vec{0},(\begin{smallmatrix}
        1&0.5\\
        0.5&1
    \end{smallmatrix}))$ for $m=100,200, \cdots,1000$ and $m=n$ (1st row) or $m=2n$ (2nd row). Three depth functions are adopted: Mahalanobis depth (1st column), Spatial depth (2nd column), and Projection depth (3rd column). $\lambda=0.1,0.3,0.5,0.7,1$. }
    \label{fig:7}
\end{figure}

\begin{figure}[H]
    \centering    \includegraphics[width=0.7\linewidth]{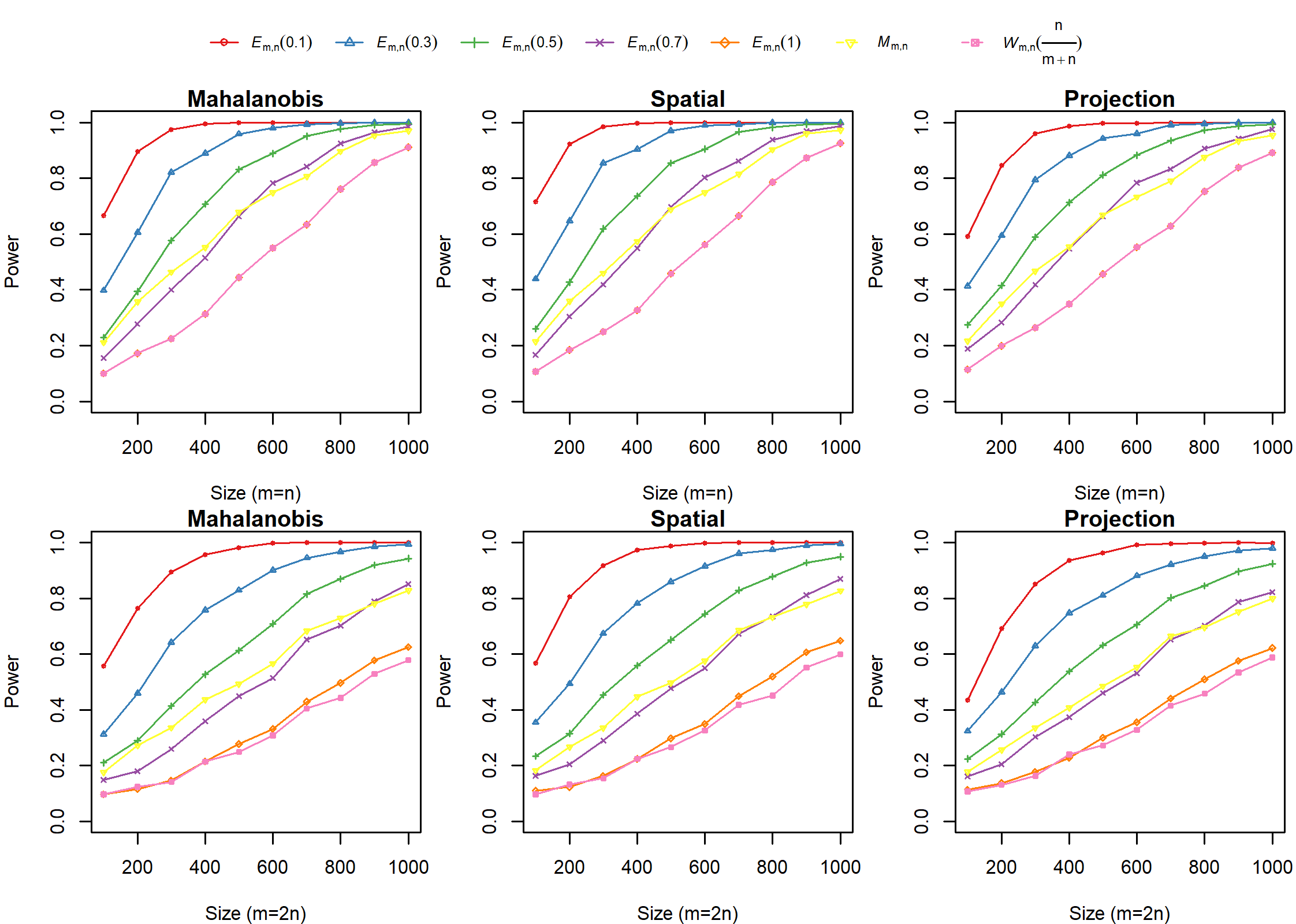}
    \caption{Power comparison under alternative hypothesis $H_a: F \sim N(\Vec{0},(\begin{smallmatrix}
        1&0\\
        0&1
    \end{smallmatrix}))$ vs $  G \sim N((\begin{smallmatrix}
        0.35\\0.35
    \end{smallmatrix}),(\begin{smallmatrix}
        1&0\\
        0&1
    \end{smallmatrix}))$ for $m=100,200, \cdots,1000$ and $m=n$ (1st row) or $m=2n$ (2nd row). Three depth functions are adopted: Mahalanobis depth (1st column), Spatial depth (2nd column), and Projection depth (3rd column). $\lambda=0.1,0.3,0.5,0.7,1$}
    \label{fig:8}
\end{figure}

\begin{figure}[H]
    \centering    \includegraphics[width=0.7\linewidth]{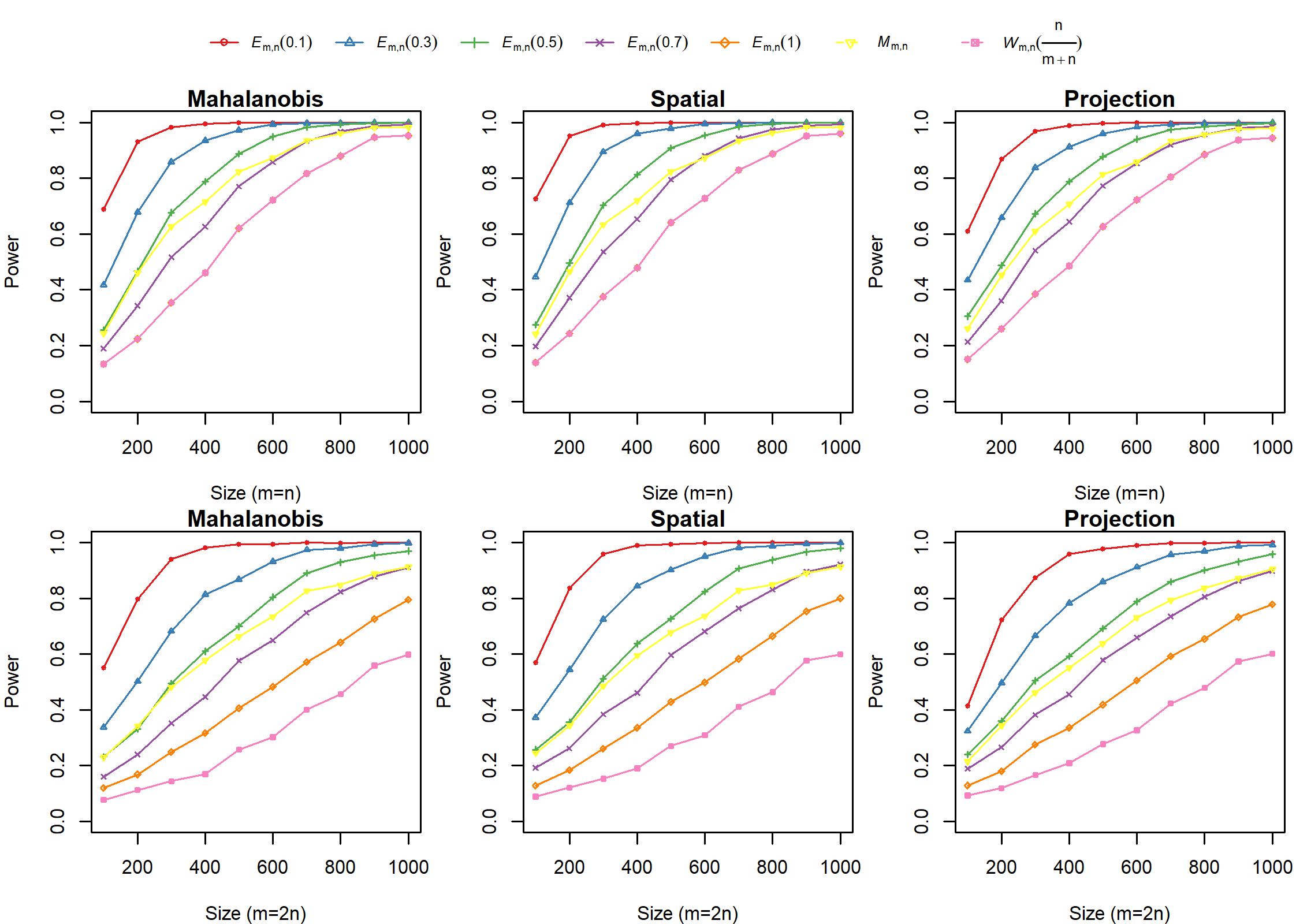}
    \caption{Power comparison under alternative hypothesis $H_a: F \sim N(\Vec{0},(\begin{smallmatrix}
        1&0\\
        0&1
    \end{smallmatrix}))$ vs $ G \sim N((\begin{smallmatrix}
        0.3\\0.3
    \end{smallmatrix}),(\begin{smallmatrix}
        1&0.4\\
        0.4&1
    \end{smallmatrix}))$ for $m=100,200, \cdots,1000$ and $m=n$ (1st row) or $m=2n$ (2nd row). Three depth functions are adopted: Mahalanobis depth (1st column), Spatial depth (2nd column), and Projection depth (3rd column). $\lambda=0.1,0.3,0.5,0.7,1$}
    \label{fig:9}
\end{figure}

All the Figures \ref{fig:7}, \ref{fig:8}, and \ref{fig:9} show that the power of $\mathcal{E}_{m,n}(\lambda)$ is increasing as the decrease of the value of $\lambda$. Compared with $M_{m,n}$, $\mathcal{E}_{m,n}(\lambda)$ outperforms when $\lambda$ is less than 0.3 (Figure \ref{fig:7}), 0.5(Figure \ref{fig:8}) and 0.5 (Figure \ref{fig:9}), respectively. To balance the type I error and power, $\lambda=0.3$ is adaptable when the sample size is bigger than 300. For larger sample size, a smaller $\lambda$ can be considered. 

\section{Analysis of Real Data Set}
\subsection{Breast Cancer Data}
Breast cancer screening helps to detect the breast cancer early and ensure a good outcome in treatment with a higher probability. \cite{breastcancer1} analyzed the data collected in routine blood analyses—notably, Glucose, Insulin, Homeostasis Model Assessment (HOMA), Leptin, Adiponectin, Resistin, Chemokine Monocyte Chemoattractant Protein 1 (MCP-1), Age and Body Mass Index (BMI)—to predict the presence of breast cancer. It is interesting to see whether the proposed test can distinguish healthy people from patients based on this data set. The blood samples were collected from 64 patients with breast cancer and 52 healthy controls at the same time of the day after an overnight fasting. 9 clinical features were observed or measured in routine blood analysis. The BMI ($kg/m^2$) was the ratio of weight and squared height. An automatic analyzer using a commercial kit determined the Serum Glucose ($mg/dL$) levels. Serum values of Resistin ($ng/mL$), MCP-1 ($pg/dL$), Leptin ($ng/mL$) and Adiponectin ($\mu g/mL$) were all assessed using commercial enzyme-linked immunosorbent assay kits from $R\&D$ System, UK, and Human MCP-1 ELISA Set, BD Biosciences Pharmingen, CA, EUA. ELISA kit used Mercodia Insulin ELISA to measure the plasma levels of Insulin ($\mu U/mL$). Insulin resistance was evaluated by the HOMA index, where $\text{HOMA}=\log\frac{If*Gf}{22.5}.$ $If$ ($\mu U/mL$) is the fasting insulin level and $Gf$ ($mmol/L$) is the fasting Glucose level.

 \begin{table*}[h]
\caption{Estimated $p$-values of ellipse statistic $\mathcal{E}_{m,n}(0.3)$, maximum statistic $M_{m,n}$, and weighted statistic $W_{m,n}(\frac{n}{m+n})$, by using three depth functions under 10,000~repetitions.}
\label{table1}
\centering
\begin{tabular}{@{}p{4cm}p{3cm}p{2.5cm}p{3.5cm}@{}}
\hline
\textbf{Depth Function} & $\mathcal{E}_{m,n}(0.3)$ & $M_{m,n}$ & $W_{m,n}(\frac{n}{m+n})$ \\ 
\hline
Mahalanobis & 0.0013 & 0.0337 & 0.0079 \\
Projection  & 0.0007 & 0.0522 & 0.0373 \\
Spatial     & 0.0010 & 0.0380 & 0.0064 \\
\hline
\end{tabular}
\end{table*}

Table \ref{table1}  shows all estimated $p$-values of rotated ellipse statistic $\mathcal{E}_{m,n}(0.3)$, maximum statistic $M_{m,n}$, and weighted statistic $W_{m,n}(\frac{n}{m+n})$, by using three depth functions under 10,000~repetitions. The estimated $p$-values are obtained by replacing the asymptotic critical value with an empirical quantile derived from simulated observations under the null hypothesis. Compared with $\alpha=0.05$, all estimated $p$-values are significant except the maximum statistic $M_{m,n}$ and weighted statistic $W_{m,n}(\frac{n}{m+n})$ for Projection depth. We note that the estimated $p$-values of rotated ellipse statistic $\mathcal{E}_{m,n}(0.3)$ are much smaller than others for all three depth functions. Compared with smaller $\alpha$, they are still significant.

\subsection{Red Wine Data}
In the second case, we consider red wine of \textit{vinho verde}, a unique band of Portugal \citep{redwine}. In the data, there are 11 features—notably, Fixed acidity, Citric acid, Residual sugar, Chlorides, Free sulfur dioxide, Total sulfur dioxide, Density, pH, Sulphates, and Alcohol, which are results of the most common physicochemical tests. The grade of the wine on a scale ranges from 0 (very bad) to 10 (excellent). In our study, we pairwise compared the wine with quality scores 4, 5, 6, and 7. The sample sizes are 53, 681, 638, and 199 respectively. Both the empirical $p$-value and asymptotic $p$-value are close to 0 for all comparison cases except the case 4 \textbf{vs} 5, and 4 \textbf{vs} 6.

\begin{table*}[ht]
\caption{Estimated $p$-values of ellipse statistic $\mathcal{E}_{m,n}(0.15)$, $\mathcal{E}_{m,n}(0.3)$, maximum statistic $M_{m,n}$, and weighted statistic $W_{m,n}\left(\frac{n}{m+n}\right)$, by using three depth functions under 10,000~repetitions.}
\label{table2}
\centering
\begin{tabular}{@{}p{3.5cm}p{1.1cm}p{1.1cm}p{1.1cm}p{1.1cm}p{1.1cm}p{1.1cm}p{1.1cm}p{1.1cm}@{}}
\hline
\textbf{Depth Function} & \multicolumn{2}{c}{$\mathcal{E}_{m,n}(0.15)$} & \multicolumn{2}{c}{$\mathcal{E}_{m,n}(0.3)$} & \multicolumn{2}{c}{$M_{m,n}$} & \multicolumn{2}{c}{$W_{m,n}\left(\frac{n}{m+n}\right)$} \\ 
& \textbf{4 vs 5} & \textbf{4 vs 6} & \textbf{4 vs 5} & \textbf{4 vs 6} & \textbf{4 vs 5} & \textbf{4 vs 6} & \textbf{4 vs 5} & \textbf{4 vs 6} \\
\hline
Mahalanobis & 0.0402 & 0      & 0.1009 & 0      & 0.7355 & 0.3407 & 0.0189 & 0.0005 \\
Projection  & 0.0007 & 0.0019 & 0.0021 & 0.0090 & 0.0087 & 0.0910 & 0.4095 & 0.3104 \\
Spatial     & 0.0401 & 0      & 0.1043 & 0      & 0.7592 & 0.3692 & 0.0144 & 0.0004 \\
\hline
\end{tabular}
\end{table*}

Table \ref{table2} shows that the ellipse test statistic has the smallest estimated $p$-value across all cases and depth functions. Compared with other test statistics, it can successfully distinguish the wines with varying quality levels, thereby supporting the Wine Rating Method. When the sample size is big enough, a smaller value of $\lambda$ can significantly improve the power of ellipse statistic. In case 4 \textbf{vs} 5, $\mathcal{E}_{m,n}(0.15)$ has a much smaller estimated $p$-value than $\mathcal{E}_{m,n}(0.3)$.

\section{Conclusions}
In this paper, we introduce a novel test statistic for assessing the homogeneity of two multivariate samples. The ellipse statistics, derived from pairwise quality indices, are shown to follow a $\chi^2_1$ asymptotic null distribution. Through simulation studies, we demonstrate the superior performance of our proposed tests. In our future work, although it presents a challenge, we would like to explore the generalization of the ellipse statistics into the multivariate multi-sample situation, which would be of interest. 

\section*{Disclosure statement}
The authors report there are no competing interests to declare.

\section*{Data Availability Statement}

The breast cancer dataset analyzed in this study is publicly available from the UCI Machine Learning Repository at 
\url{https://archive.ics.uci.edu/ml/datasets/Breast+Cancer+Coimbra}. The wine quality dataset analyzed in this study is also publicly available from the UCI Machine Learning Repository at 
\url{https://archive.ics.uci.edu/ml/datasets/Wine+Quality}.

\section*{Funding}
Dr. Shi's work was supported by the Natural Sciences and Engineering Research Council of Canada under Grant RGPIN-2022-03264, the NSERC Alliance International Catalyst Grant ALLRP 590341-23, and the University of British Columbia Okanagan (UBC-O) Vice Principal Research in collaboration with UBC-O Irving K. Barber Faculty of Science. Dr. Fu’s research was supported by NSERC Discovery Grant RGPIN 2018-05846. Dr. Chen’s research was supported by NSERC Discovery Grant
RGPIN 2022-04519.

\begin{appendix}
\section{Proof of Theorem 1}\label{appa}
\cite{variations} shows that under A1-A4 and null hypothesis $H_0:F=G$, for $m,n \to \infty,$ 
$$
\frac{m}{m+n} \to \tau \textit{ for some } 0 < \tau <1,
$$ we have 
\begin{equation*}
\sqrt{\frac{12mn}{m+n}}
\begin{pmatrix}
\mathcal{Q}(F_m,G_n)-\frac{1}{2}\\
\mathcal{Q}(G_n,F_m) -\frac{1}{2}
\end{pmatrix}
\xrightarrow{\enskip D \enskip} 
\begin{pmatrix}
    1\\
    -1
\end{pmatrix}z, \text{ where } z \sim 
N(0,1).    
\end{equation*} 
According to the Continuous Mapping Theorem,
\begin{eqnarray*}
    \mathcal{R}(\lambda,\theta,F_m,G_n)    
    &=&    
    \frac{12mn}{m+n}
    \vec{\mathcal{Q}}^T
    \begin{pmatrix}
    \cos^2{\theta}+\lambda \sin^2{\theta}& (1-\lambda)\sin{\theta}\cos{\theta}\\
    (1-\lambda)\sin{\theta}\cos{\theta} & \sin^2{\theta}+\lambda \cos^2{\theta}
    \end{pmatrix}
    \vec{\mathcal{Q}}\\    
    &\xrightarrow{\enskip D \enskip}&    
    \Big[\begin{pmatrix}
    1\\-1
    \end{pmatrix}z\Big]^T\begin{pmatrix}
    \cos^2{\theta}+\lambda \sin^2{\theta}& (1-\lambda)\sin{\theta}\cos{\theta}\\
    (1-\lambda)\sin{\theta}\cos{\theta} & \sin^2{\theta}+\lambda \cos^2{\theta}
    \end{pmatrix}
    \Big[\begin{pmatrix}
    1\\-1
    \end{pmatrix}z\Big]\\    
    &=& 
    z^T\begin{pmatrix}    
    1&-1
    \end{pmatrix}
    \begin{pmatrix}
    \cos^2{\theta}+\lambda \sin^2{\theta}& (1-\lambda)\sin{\theta}\cos{\theta}\\
    (1-\lambda)\sin{\theta}\cos{\theta} & \sin^2{\theta}+\lambda \cos^2{\theta}
    \end{pmatrix}
    \begin{pmatrix}
    1\\-1
    \end{pmatrix}z\\    
    &=&    
    [1+\lambda-(1-\lambda)\sin{2\theta}]z^Tz\\    
    &\sim&    
    [1+\lambda-(1-\lambda)\sin{2\theta}]\chi^2_1
\end{eqnarray*}
where $\vec{\mathcal{Q}}=\begin{pmatrix}
    \mathcal{Q}(F_m,G_n)-\frac{1}{2}\\
    \mathcal{Q}(G_n,F_m)-\frac{1}{2}
\end{pmatrix}$. 

\section{Distributions Specified for Alternative Hypothesis Scenarios}\label{appb}
$F = 0.8N(\Vec{0},\mathbf{I})+0.2N((\begin{smallmatrix}
        15\\
        15
    \end{smallmatrix}),\mathbf{I})$
\begin{itemize}
    \item Green points:
    $$G=0.6N((\begin{smallmatrix}
        5\\
        5
    \end{smallmatrix}),\mathbf{I})+0.4N((\begin{smallmatrix}
        -2\\
        -2
    \end{smallmatrix}),\mathbf{I})$$
    \item Blue points:
    $$G=0.8N((\begin{smallmatrix}
        4\\
        4
    \end{smallmatrix}),(\begin{smallmatrix}
        0.9&0\\
        0&0.9
    \end{smallmatrix}))+0.2N((\begin{smallmatrix}
        -5\\
        -5
    \end{smallmatrix}),(\begin{smallmatrix}
        0.9&0\\
        0&0.9
    \end{smallmatrix}))$$
    \item Yellow points:
    $$G=0.8N((\begin{smallmatrix}
        0\\
        0
    \end{smallmatrix}),(\begin{smallmatrix}
        0.8&0\\
        0&0.8
    \end{smallmatrix}))+0.2N((\begin{smallmatrix}
        -12\\
        -12
    \end{smallmatrix}),(\begin{smallmatrix}
        0.7&0\\
        0&0.7
    \end{smallmatrix}))$$
    \item Brown points:
    $$G=0.8N((\begin{smallmatrix}
        3\\
        3
    \end{smallmatrix}),\mathbf{I})+0.2N((\begin{smallmatrix}
        -1\\
        -1
    \end{smallmatrix}),(\begin{smallmatrix}
        2&0\\
        0&2
    \end{smallmatrix}))$$
    \item Red points:
    $$G=0.8N((\begin{smallmatrix}
        7\\
        7
    \end{smallmatrix}),\mathbf{I})+0.2N((\begin{smallmatrix}
        -8\\
        -8
    \end{smallmatrix}),\mathbf{I})$$
\end{itemize}    
\end{appendix}

\bibliography{references}

@incollection{liu1992data,
  author    = {Liu, R.Y.},
  title     = {Data depth and multivariate rank tests},
  booktitle = {L1-Statistical Analysis and Related Methods},
  pages     = {279--294},
  publisher = {North-Holland},
  address   = {Amsterdam},
  year      = {1992}
}

@article{liu1993,
  author  = {Liu, R.Y. and Singh, K.},
  title   = {A quality index based on data depth and multivariate rank tests},
  journal = {Journal of the American Statistical Association},
  volume  = {88},
  number  = {421},
  pages   = {252--260},
  year    = {1993}
}

@article{shi1,
  author  = {Shi, X. and Zhang, Y. and Fu, Y.},
  title   = {Two-sample tests based on data depth},
  journal = {Entropy},
  volume  = {25},
  number  = {2},
  pages   = {238},
  year    = {2023},
  note    = {Article 238}
}

@article{zuo2006limiting,
  author  = {Zuo, Y. and He, X.},
  title   = {On the limiting distributions of multivariate depth-based rank sum statistics and related tests},
  journal = {Annals of Statistics},
  volume  = {34},
  number  = {6},
  pages   = {2879--2896},
  year    = {2006}
}

@book{finance,
  author    = {Rachev, S.T. and Mittnik, S. and Fabozzi, F.J. and Focardi, S.M.},
  title     = {Financial Econometrics: From Basics to Advanced Modeling Techniques},
  publisher = {Wiley},
  address   = {New York},
  year      = {2007}
}

@article{ecology,
  author  = {Neuh{\"a}user, M. and Ruxton, G.D.},
  title   = {Distribution-free two-sample comparisons in the case of heterogeneous variances},
  journal = {Behavioral Ecology and Sociobiology},
  volume  = {63},
  pages   = {617--623},
  year    = {2009}
}

@article{microbiology,
  author  = {Banerjee, K. and Zhao, N. and Srinivasan, A. and Xue, L. and Hicks, S.D. 
             and Middleton, F.A. and Wu, F.A. and Zhan, X.},
  title   = {An adaptive multivariate two-sample test with application to microbiome differential abundance analysis},
  journal = {Frontiers in Genetics},
  volume  = {10},
  pages   = {447114},
  year    = {2019},
  note    = {Article 447114}
}

@article{zuo2000general,
  author  = {Zuo, Y. and Serfling, R.},
  title   = {General notions of statistical depth function},
  journal = {Annals of Statistics},
  volume  = {28},
  pages   = {461--482},
  year    = {2000}
}

@article{cramertest,
  author  = {Baringhaus, L. and Franz, C.},
  title   = {On a new multivariate two-sample test},
  journal = {Journal of Multivariate Analysis},
  volume  = {88},
  pages   = {190--206},
  year    = {2004}
}

@article{energytest,
  author  = {Sz{\'e}kely, G. and Rizzo, M.},
  title   = {Testing for equal distributions in high dimension},
  journal = {InterStat},
  volume  = {5},
  year    = {2004}
}

@article{variations,
  author  = {Gnettner, F. and Kirch, C. and Nieto-Reyes, A.},
  title   = {Symmetrisation of a class of two-sample tests by mutually considering depth ranks including functional spaces},
  journal = {Electronic Journal of Statistics},
  volume  = {18},
  number  = {2},
  pages   = {3021--3106},
  year    = {2024}
}

@article{kernel1,
  author  = {Gretton, A. and Borgwardt, K.M. and Rasch, M.J. and Sch{\"o}lkopf, B. and Smola, A.},
  title   = {A kernel two-sample test},
  journal = {Journal of Machine Learning Research},
  volume  = {13},
  number  = {1},
  pages   = {723--773},
  year    = {2012}
}

@article{maxmean,
  author  = {Zhang, J.-T. and Smaga, {\L}.},
  title   = {Two-sample test for equal distributions in separate metric space: New maximum mean discrepancy based approaches},
  journal = {Electronic Journal of Statistics},
  volume  = {16},
  number  = {2},
  pages   = {4090--4132},
  year    = {2022}
}

@article{family,
  author  = {Ramdas, A. and Trillos, N.G. and Cuturi, M.},
  title   = {On Wasserstein two-sample testing and related families of nonparametric tests},
  journal = {Entropy},
  volume  = {19},
  number  = {2},
  pages   = {47},
  year    = {2017},
  note    = {Article 47}
}

@article{chara1,
  author  = {Liu, R. and Parelius, J. and Singh, K.},
  title   = {Multivariate analysis by data depth: Descriptive statistics, graphics and inference},
  journal = {Annals of Statistics},
  volume  = {27},
  year    = {1999}
}

@article{liu2006,
author = {Liu, Regina and Singh, Kesar},
year = {2006},
month = {11},
pages = {17-35},
title = {Rank tests for multivariate scale difference based on data depth},
volume = {72},
journal = {DIMACS: Series in Discrete Mathematics and Theoretical Computer Science},
doi = {10.1090/dimacs/072/02}
}

@article{mod3,
  author  = {Chenouri, S. and Small, C.G. and Farrar, T.J.},
  title   = {Data depth-based nonparametric scale tests},
  journal = {Canadian Journal of Statistics},
  volume  = {39},
  number  = {2},
  pages   = {356--369},
  year    = {2011}
}

@article{hotel1,
  author  = {Lawley, D.N.},
  title   = {A generalization of Fisher's Z test},
  journal = {Biometrika},
  volume  = {30},
  number  = {1/2},
  pages   = {180--187},
  year    = {1938}
}

@inproceedings{hotel2,
  author    = {Hotelling, H.},
  title     = {A generalized T test and measure of multivariate dispersion},
  booktitle = {Proceedings of the Second Berkeley Symposium on Mathematical Statistics and Probability},
  volume    = {2},
  pages     = {23--42},
  year      = {1951}
}

@incollection{spatial1,
  author    = {Serfling, R.},
  title     = {A depth function and a scale curve based on spatial quantiles},
  booktitle = {Statistical Data Analysis Based on the L1-Norm and Related Methods},
  pages     = {25--38},
  publisher = {Birkh{\"a}user},
  address   = {Basel},
  year      = {2002}
}

@article{spatial2,
  author  = {Vardi, Y. and Zhang, C.H.},
  title   = {The multivariate $L_1$-median and associated data depth},
  journal = {Proceedings of the National Academy of Sciences of the USA},
  volume  = {97},
  number  = {4},
  pages   = {1423--1426},
  year    = {2000}
}

@article{spatial3,
  author  = {Pokotylo, O. and Mozharovskyi, P. and Dyckerhoff, R.},
  title   = {Depth and depth-based classification with R package ddalpha},
  journal = {Journal of Statistical Software},
  volume  = {91},
  number  = {5},
  pages   = {1--46},
  year    = {2019}
}

@article{ansari1960rank,
  author  = {Ansari, A.R. and Bradley, R.A.},
  title   = {Rank-sum tests for dispersions},
  journal = {Annals of Mathematical Statistics},
  volume  = {31},
  pages   = {1174--1189},
  year    = {1960}
}

@article{siegel1960nonparametric,
  author  = {Siegel, S. and Tukey, J.W.},
  title   = {A nonparametric sum of squares test for homogeneity of variance in several populations},
  journal = {Journal of the American Statistical Association},
  volume  = {55},
  number  = {290},
  pages   = {129--143},
  year    = {1960}
}

@article{breastcancer1,
  author  = {Patr{\'\i}cio, M. and Pereira, J. and Cris{\'o}stomo, J. and Matafome, P. and Gomes, M. and Sei{\c{c}}a, R. and Caramelo, F.},
  title   = {Using resistin, glucose, age and BMI to predict the presence of breast cancer},
  journal = {BMC Cancer},
  volume  = {18},
  pages   = {1--8},
  year    = {2018}
}

@article{redwine,
  author  = {Cortez, P. and Cerdeira, A.L. and Almeida, F. and Matos, T. and Reis, J.},
  title   = {Modeling wine preferences by data mining from physicochemical properties},
  journal = {Decision Support Systems},
  volume  = {47},
  pages   = {547--553},
  year    = {2009}
}

\end{document}